\documentclass[twocolumn,english,prl]{revtex4}
\usepackage[T1]{fontenc}
\usepackage[utf8]{inputenc}
\usepackage{color}
\usepackage{float}
\usepackage{amsmath}
\usepackage{graphicx}
\usepackage{amssymb}

\makeatletter

\providecommand{\tabularnewline}{\\}

\@ifundefined{textcolor}{}
{%
 \definecolor{BLACK}{gray}{0}
 \definecolor{WHITE}{gray}{1}
 \definecolor{RED}{rgb}{1,0,0}
 \definecolor{GREEN}{rgb}{0,1,0}
 \definecolor{BLUE}{rgb}{0,0,1}
 \definecolor{CYAN}{cmyk}{1,0,0,0}
 \definecolor{MAGENTA}{cmyk}{0,1,0,0}
 \definecolor{YELLOW}{cmyk}{0,0,1,0}
 }

\makeatother

\usepackage{babel}

\begin{document}

\title{Demonstrating quantum speed-up in a superconducting two-qubit processor }

\author{A. Dewes$^{1}$, R. Lauro$^{1},$ F.R. Ong$^{1},$ V. Schmitt$^{1}$,
P. Milman$^{2,3}$, P. Bertet$^{1}$, D. Vion$^{1}$, and D. Esteve$^{1}$}

\affiliation{$^{1}$Service de Physique de l'Etat Condens{é}/IRAMIS/DSM (CNRS
URA 2464), CEA Saclay, 91191 Gif-sur-Yvette, France}

\affiliation{$^{2}$Laboratoire Matériaux et Phénomènes Quantiques, Université
Paris Diderot, 10 rue Alice Domon et Léonie Duquet, 75205 Paris, France, }

\affiliation{$^{3}$Univ. Paris-Sud 11, Institut de Sciences Moléculaires d'Orsay
(CNRS), 91405 Orsay, France}

\date{\today}
\begin{abstract}
We operate a superconducting quantum processor consisting of two tunable
transmon qubits coupled by a swapping interaction, and equipped with
non destructive single-shot readout of the two qubits. With this processor,
we run the Grover search algorithm among four objects and find that
the correct answer is retrieved after a single run with a success
probability between $0.52$ and $0.67$, significantly larger than
the $0.25$ achieved with a classical algorithm. This constitutes
a proof-of-concept for the quantum speed-up of electrical quantum
processors. 
\end{abstract}
\maketitle
\begin{figure}
\includegraphics[width=8cm]{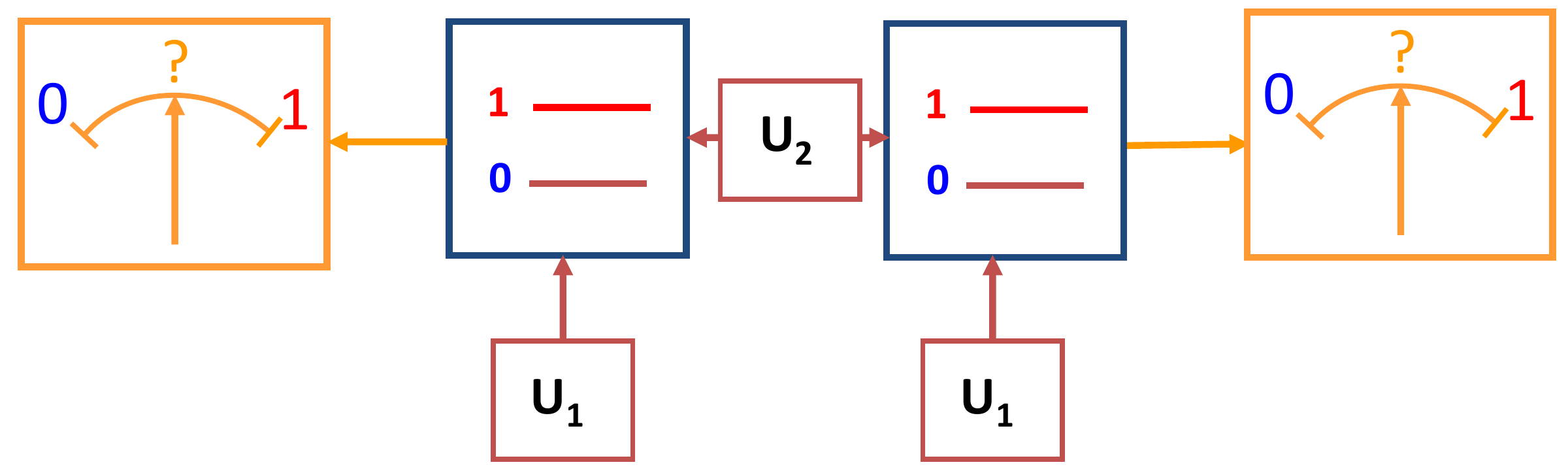}

\caption{\label{fig:blueprint}Schematic blueprint of a quantum processor based
on quantum gates, represented here in the two-qubit case relevant
for our experiment. A quantum processor consists of a qubit register
that can perform any unitary evolution needed by an algorithm under
the effect of a universal set of quantum gates (single qubit gate
$U_{1}$ , two-qubit gate $U_{2}$). Ideally, all the qubits may be
read projectively, and may be reset.}

\end{figure}

The proposition of quantum algorithms \cite{search,Shor,NielsenChuang}
that perform useful computational tasks more efficiently than classical
algorithms has motivated the realization of physical systems \cite{QC}
able to implement them and to demonstrate quantum speed-up. The versatility
and the potential scalability of electrical circuits make them very
appealing for implementing a quantum processor built as sketched in~Fig.
\ref{fig:blueprint}. Ideally, a quantum processor consists of a scalable
set of quantum bits that\textcolor{black}{{} can be efficiently reset,
that can follow any unitary evolution needed by an algorithm using
a universal set of single and two qubit gates, and that can be read
projectively} \cite{divincenzo}. The non-unitary projective readout
operations can be performed at various stages of an algorithm, and
in any case at the end in order to get the final outcome. Quantum
processors based on superconducting qubits have already been operated,
but they fail to meet the above criteria in different aspects. With
the transmon qubit \cite{Transmon Koch,TransmonSchreier} derived
from the Cooper pair box \cite{box}, two simple quantum algorithms,
namely the Deutsch-Jozsa algorithm \cite{DeutschJozsa} and the Grover
search algorithm \cite{search}, were demonstrated in a two qubit
processor with the coupling between the qubits mediated by a cavity
also used for readout \cite{dicarlo}. In this circuit, the qubits
are not read independently, but the value of a single collective variable
is determined from the cavity transmission measured over a large number
of repeated sequences. By applying suitable qubit rotations prior
to this measurement, the density matrix of the two-qubit register
was inferred at different steps of the algorithm, and found in good
agreement with the predicted one. Demonstrating quantum speed-up is
however more demanding than measuring a collective qubit variable
since it requests to obtain an outcome after a single run, i.e. to
perform the single-shot readout of the qubit register. Up to now,
single-shot readout in superconducting processors has been achieved
only for phase qubits \cite{yamamoto,mariantoni}. In a two phase-qubit
processor equipped with single-shot but destructive readout of the
qubits, the Deutsch-Jozsa algorithm \cite{DeutschJozsa} was demonstrated
in Ref.$\,$\cite{yamamoto} with a success probability of order $0.7$
in a single run, to be compared to $0.5$ for a classical algorithm. 

Since the Deutsch-Jozsa classification algorithm is not directly related
to any practical situation, demonstrating quantum speed-up for more
useful algorithms in an electrical processor designed along the blueprint
of Fig.~\ref{fig:blueprint} is an important goal. In this work,
we operate a new two transmon-qubit processor \cite{processor} that
comes closer to the ideal scheme than those previously mentioned,
and we run the Grover search algorithm among four objects. Since,
in this case, the algorithm ideally yields the answer after one algorithm
step, its success probability after a single run provides a simple
benchmark. We find that our processor yields the correct answer at
each run with a success probability that ranges between $0.52$ and
$0.67$, whereas a single step classical algorithm using a random
query would yield the correct answer with probability $0.25$. 

\begin{figure}
\includegraphics[width=8.5cm]{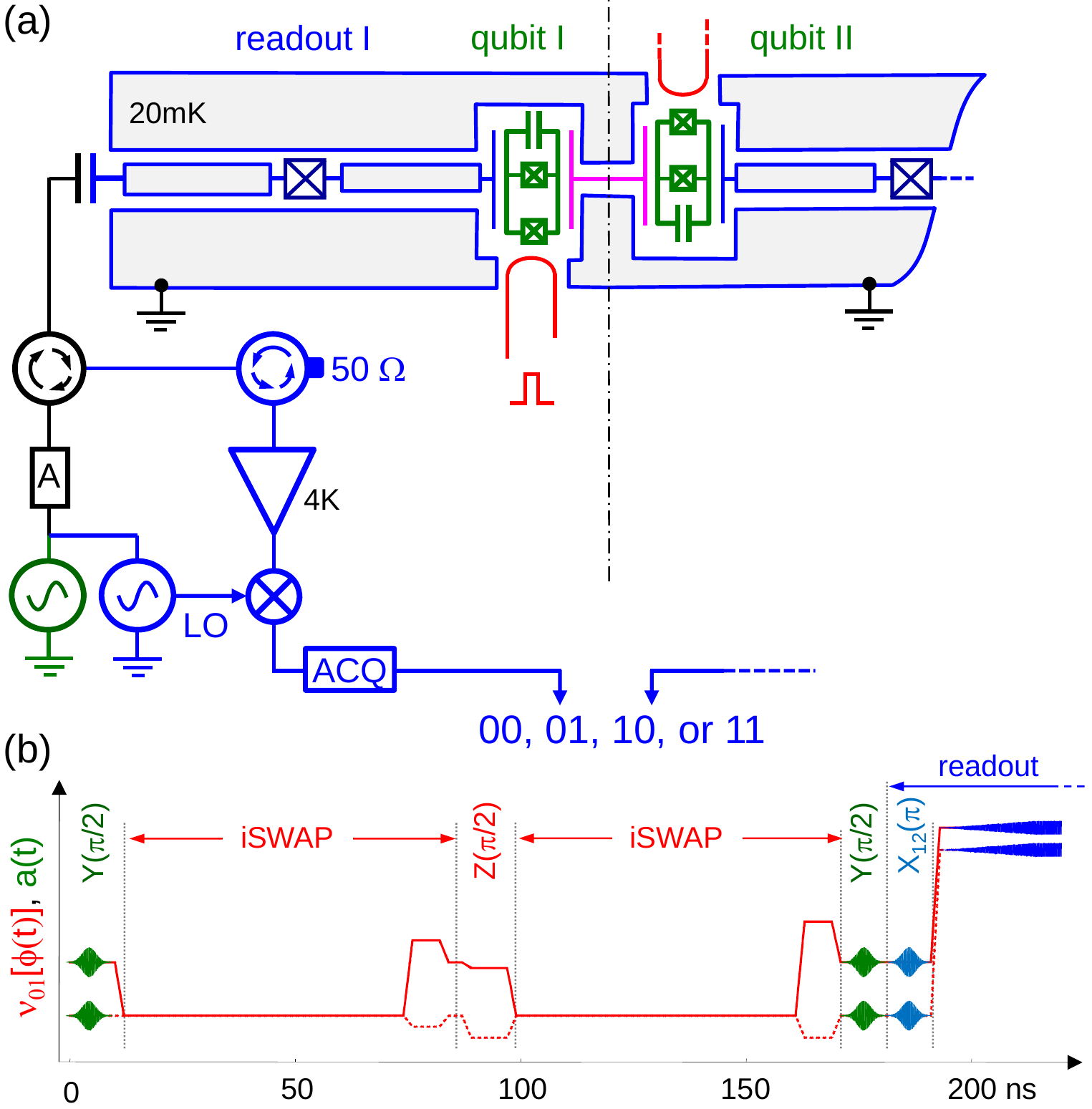} \caption{\label{fig:circuittomo} Electrical scheme of the two qubit circuit
operated and typical sequence during processor operation. (a) Two
capacitively coupled transmon qubits (green) have tunable frequencies
controlled by the flux induced in their SQUID loop by a local current
line (in red). The coupling capacitance (in magenta) yields a swapping
evolution between the qubits when on resonance. Each transmon is embedded
in a non-linear resonator used for single-shot readout. Each reflected
readout pulse is routed to a cryogenic amplifier through circulators,
homodyned at room temperature and acquired digitally, which yields
a two-bit outcome. (b) Typical operation of the processor showing
the resonant microwave pulses $a(t)$ applied to the qubits (green)
and to the readouts (blue), on top of the DC pulses (red lines) that
vary the transition frequencies of qubit $I$ (solid) and $II$ (dashed).
With the qubits tuned at a first working point for single qubit gates,
resonant pulses are applied for performing X and Y rotations, as well
as small flux pulses for Z rotations; qubits are then moved to the
interaction point for two-qubit gate operations. Such sequences can
be combined as needed by the algorithm. Qubits are then moved to their
initial working points for applying tomography pulses as well as a
$\left|1\right\rangle \rightarrow\left|2\right\rangle $ pulse $X_{12}(\pi)$
to increase the fidelity of the forthcoming readout. Finally, they
are moved to better readout points and read. }
\end{figure}

The scheme and the operation mode of our processor is shown in Fig.~\ref{fig:circuittomo}.
Two tunable transmon qubits coupled by a fixed capacitor, are embedded
in two identical control and readout sub-circuits. The Hamiltonian
of the two qubits $\left\{ I,II\right\} $ is $H/h=\left(-\nu_{\mathrm{I}}\sigma_{\mathrm{z}}^{\mathrm{I}}-\nu_{\mathrm{II}}\sigma_{\mathrm{z}}^{\mathrm{II}}+2g\sigma_{\mathrm{y}}^{\mathrm{I}}\sigma_{\mathrm{y}}^{\mathrm{II}}\right)/2$,
where $\sigma_{\mathrm{x,y,z}}$ are the Pauli operators, $\nu_{\mathrm{I,II}}$
the qubit frequency controlled by the flux applied to each transmon
SQUID loop with a fast ($0.5\,\mathrm{GHz}$ bandwidth) local current
line, and $g=4.6\:\mathrm{MHz\ll\nu_{I,II}}$ the coupling frequency
controlled by the coupling capacitance. The achieved frequency control
allows us to place the two transmons on resonance during times precise
enough for performing the universal two-qubit gate \textcolor{black}{$\sqrt{iSWAP}$}
\cite{processor} and the exchange gate\textcolor{black}{{} $iSWAP$}
used in this work. The qubit frequencies are tuned to different values
for single qubit manipulation, two-qubit gate operation, and readout.
The readout is independently and simultaneously performed for each
qubit using the single-shot method of Ref.$\,$\cite{mallet}. It
is based on the dynamical transition of a non-linear resonator \cite{siddiqi,metcalfe}
that maps the quantum state of each transmon to the bifurcated/non
bifurcated state of its resonator,  which yields a binary outcome
for each qubit. This readout method is potentially non-destructive,
but its non-destructive character is presently limited by relaxation
during the readout pulse. In order to further improve the readout
fidelity, we resort to a shelving method that exploits the second
excited state of the transmon. For this purpose, a microwave pulse
that induces a transition from the state $\left|1\right\rangle $
towards the second excited state $\left|2\right\rangle $ of the transmon
is applied just before the readout pulse as demonstrated in Ref.$\,$\cite{mallet}.
This variant does not alter the non-destructive aspect of the readout
method since an extra pulse bringing state $\left|2\right\rangle $
back to state $\left|1\right\rangle $ could be applied after readout.
Although the readout contrast achieved with this shelving method and
with optimized microwave pulses reaches $0.88$ and $0.89$ for the
two qubits respectively, the values achieved at working points suitable
for processor operation are lower and equal to $0.84$ and $0.83$.
The readout outcome probabilities for all input states of the two-qubit
register are given in the Supplementary Information S4, with a discussion
of the error sources.

In order to characterize the evolution of the quantum register during
the algorithm, we determine its density matrix by state tomography.
For this purpose, we measure the expectation values of the extended
Pauli set of operators $\left\{ \sigma_{\mathrm{x}}I,..,\sigma_{\mathrm{z}}\sigma_{\mathrm{z}}\right\} $
by applying the suitable rotations just before readout and by averaging
typically $10^{4}\:$times. Note that the readout errors, which can
be well-characterized, are corrected when determining the expectation
value of the Pauli set, and thus do not contribute to tomography errors
as explained in Ref.$\,$\cite{processor}. The density matrix $\rho$
is then taken as the acceptable positive-semidefinite matrix that,
according to the Hilbert-Schmidt distance, is the closest to the possibly
non physical one derived from the measurement set. In order to characterize
the fidelity of the algorithm at all steps, we use the state fidelity
$F=\left\langle \psi\left|\rho\right|\psi\right\rangle $ with $\left|\psi\right\rangle $
the ideal quantum state at the step considered; $F$ is in this case
the probability for the qubit register to be in state $\left|\psi\right\rangle $. 

\begin{figure}[H]
\includegraphics[width=8cm]{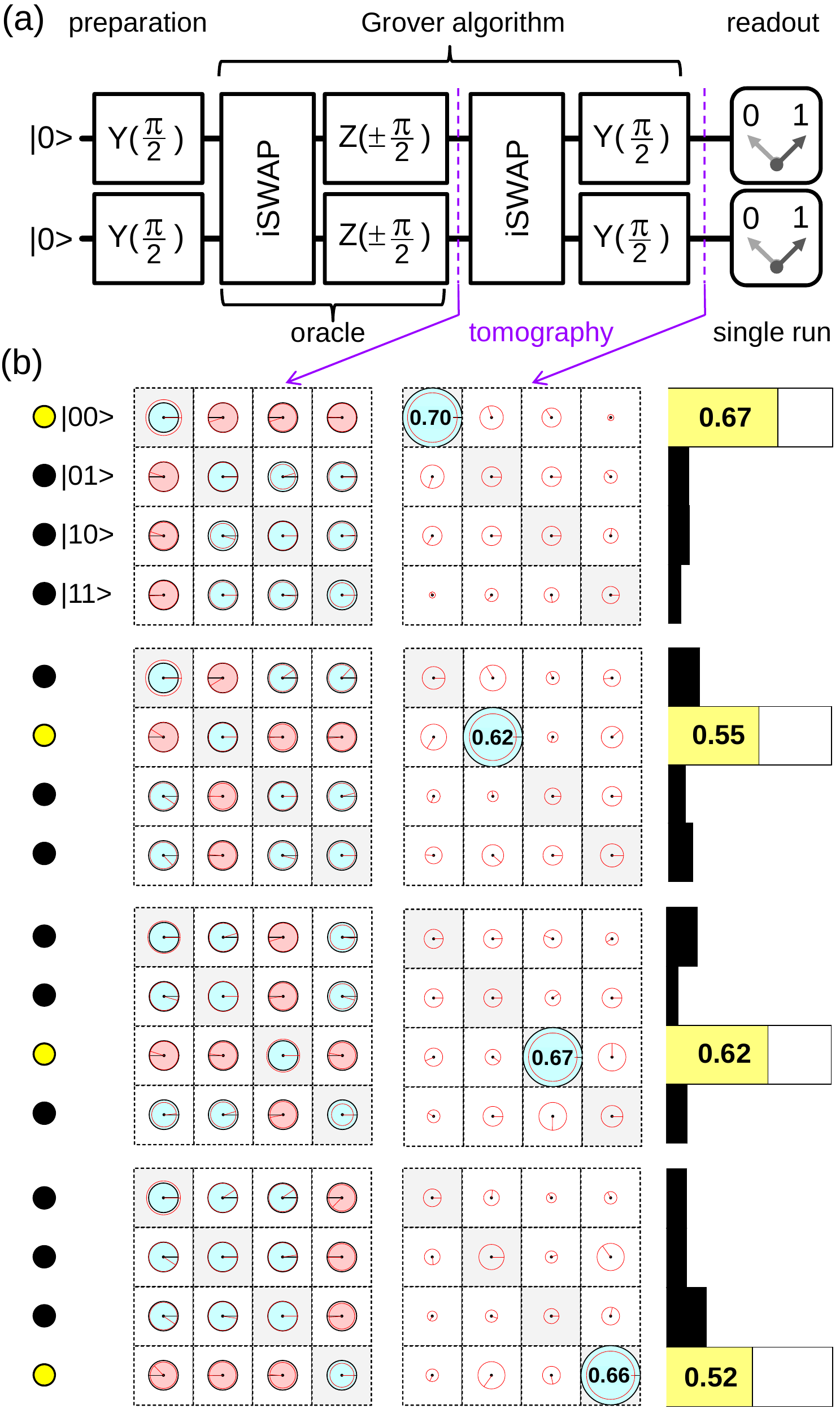} \caption{\label{fig:operation} (a) Experimental sequence used for implementing
the Grover search algorithm on four objects. First, $Y(\pi/2)$ rotations
are applied to produce the superposition $\left|\phi\right\rangle =(1/2){\textstyle \sum_{u,v}\left|uv\right\rangle }$
of all basis states; then one of the four possible oracle (corresponding
to the four sign combinations of the Z rotations) is applied. The
tagged state is then decoded in all cases using a $iSWAP$ operation
followed by $Y(\pi/2)$ rotations. (b) State tomography at two steps
of the algorithm and success probability after a single run. The yellow
dot on the left marks the basis state tagged by each oracle operator
used. After applying the oracle, the information on the tagged state
is encoded in the phase of six particular elements of the density
matrix $\rho$. After decoding, the tagged state should be the only
matrix element present in $\rho$. The amplitude of each matrix element
is represented by a disk (black for the ideal density matrix, red
for the measured one) and its phase by a radius (as well as a filling
color for the ideal matrice). The probability distribution of the
single-run readout outcomes is indicated on the right (yellow box
for the correct answer, filled dark boxes for the wrong ones).}
\end{figure}

The Grover search algorithm \cite{search} consists in retrieving
a particular basis state in a Hilbert space of size $N$ using a function
able to discriminate it from the other ones. This function is used
to build an oracle operator that tags the searched state. Starting
from the superposition $\left|\phi\right\rangle $ of all register
states, a unitary sequence that incorporates the oracle operator is
repeated about $\sqrt{N}$ times, and eventually yields the searched
state with a high probability. The implementation of Grover's algorithm
in a two-qubit Hilbert space often proceeds in a simpler way \cite{grov_Chuang,grov_Fourier_Optics,grov_Jones,grov_optics,grov_Rydberg,grov_trapped_ions}
since the result is obtained with certainty after a single algorithm
step. The algorithm then consists of an encoding sequence depending
on the searched state, followed by a universal decoding sequence that
retrieves it. Grover's algorithm thus provides a simple benchmark
for two-qubit processors. Its implementation with our quantum processor
is shown in \textcolor{black}{Fig.~\ref{fig:operation}(a). First,
the superposed state $\left|\phi\right\rangle $ is obtained by applying
$\pi/2$ rotations around the $Y$ axis for the two qubits. The oracle
operator $O_{uv}$ tagging the two-qubit state $\left|uv\right\rangle \equiv\left|u\right\rangle _{\mathrm{I}}\otimes\left|v\right\rangle _{\mathrm{II}}$
to be searched is then applied to state $\left|\phi\right\rangle $.
Each $O_{\mathrm{uv}}$ consists of a $\mathit{\mathrm{\mathit{iSWAP}}}$
gate followed by a $Z(\pm\pi/2)$ rotation on each qubit, with the
four possible sign combinations $(-,-)$, $(+,-)$ , $(-,+)$ , and
$(+,+)$ corresponding to $uv=00$, $01$, $10$, and $11$, respectively.
In the algorithm we use, as in Ref.$\,$\cite{dicarlo}}, the encoding
is a phase encoding. When applied to $\left|\phi\right\rangle $,
each oracle operator inverts the sign of the component corresponding
to the state it tags, respectively to the other ones. The density
matrix after applying the oracle ideally takes a simple form: the
amplitude of all coefficients is $1/4$, and the phase of an element
$\rho_{\mathrm{rs}}$ is $\varphi_{\mathrm{rs}}=\pi(\delta_{\mathrm{rt}}+\delta_{\mathrm{st}})$,
where $t$ corresponds to the state tagged by the oracle operator.
\textcolor{black}{{} The state tomography performed after applying the
oracle, shown in Fig.~\ref{fig:operation}(b), is in good agreement
with this prediction. More quantitatively, we find that after having
applied the oracle operator, the intermediate fidelity is $F_{\mathrm{int}}=0.87$,
$0.80$, $0.84$, and $0.82$, respectively. The last part of the
algorithm consists in transforming the obtained state in the searched
state irrespectively of it, or equivalently to transform the phase
information distributed over the elements of the density matrix in
a weight information with the whole weight on the searched state.
This operation is readily performed by applying an $\mathrm{\mathit{iSWAP}}$
gate followed by $Y(\pi/2)$ rotations on each qubit. We find that
the fidelity of the density matrix at the end of the algorithm is
$F_{\mathrm{final}}=0.70$, $0.62$, $0.67$, and $0.66$ respectively.
We explain both $F_{\mathrm{int}}$ and $F_{\mathrm{final}}$ by gate
errors at a 2\% level, by errors in the tomography pulses at a 2\%
level, as well as by decoherence during the whole experimental sequence
(at the coupling point, relaxation times are $T_{1}^{\mathrm{I}}\simeq450\,\mathrm{ns}$
and $T_{1}^{\mathrm{II}}\simeq500\,\mathrm{ns}$, and the effective
dephasing times $T_{\varphi}^{\mathrm{I}}\simeq T_{\varphi}^{\mathrm{II}}\simeq2\,\mathrm{\mathrm{\mu s}}$
\cite{processor}). $F_{\mathrm{final}}$ is thus approximately the
success probability one would obtain assuming no errors when reading
the qubit register at the end of the algorithm.}

We now consider the success probability obtained after a single run
(with no tomography pulses), which probes the quantum speed-up actually
achieved by the processor. We find (see Fig.~\ref{fig:operation})
that our processor does yield the correct answer with a success probability
$P_{\mathrm{S}}=0.67$, $0.55$, $0.62$, and $0.52$ for the four
basis states, which is smaller than the density matrix fidelity $F_{\mathrm{final}}$.
One notices that the difference between $F_{\mathrm{final}}$ and
$P_{\mathrm{S}}$, mostly due to readout errors, slightly depends
on the searched state: the larger the energy of the searched state,
the larger the difference. This dependence is well explained by the
effect of relaxation during the readout pulse, which is the main error
source at readout, the second one being readout crosstalk. One also
notices that the outcome errors are distributed over all the wrong
answers. To summarize, the error in the outcome of Grover's algorithm
originate both from small unitary errors accumulated during the algorithm,
and from decoherence during the whole sequence, in particular during
the final readout. 

\textcolor{black}{We now discuss the significance of the obtained
results in terms of quantum information processing. The achieved success
probability is smaller than the theoretically achievable value $1$,
but nevertheless sizeably larger than the value of $0.25$ obtained
by running once the classical algorithm that consists in making a
random trial. From the point of view of a user that would search which
unknown oracle picked at random has been given to him, the fidelity
of the algorithm outcome is $F_{o}=0.57$, $0.63$, $0.57$, and $0.59$
for the $00$, $01$, $10$, and $11$ outcomes respectively, as explained
in the Supplementary Information S5. Despite the presence of errors,
this result demonstrates the quantum speed-up for Grover's algorithm
when searching in a Hilbert space with small size $N=4$. Demonstrating
the $\sqrt{N}$ speed-up for Grover's algorithm in larger Hilbert
spaces requires a qubit architecture more scalable than the present
one, which presently is a major challenge in the field. }

\textcolor{black}{In conclusion, we have demonstrated the operation
of the Grover search algorithm in a superconducting two-qubit processor
with single-shot non destructive readout. This result indicates that
the quantum speed-up expected from quantum algorithms is within reach
of superconducting quantum bit processors.}

\section*{Supplementary material }

\paragraph{S1. Sample preparation\protect \\
}

The sample is fabricated on a silicon chip oxidized over 50 nm. A
150 nm thick niobium layer is first deposited by magnetron sputtering
and then dry-etched in a $SF_{6}$ plasma to pattern the readout resonators,
the current lines for frequency tuning, and their ports. Finally,
the transmon qubit, the coupling capacitance and the Josephson junctions
of the resonators are fabricated by double-angle evaporation of aluminum
through a shadow mask patterned by e-beam lithography. The first layer
of aluminum is oxidized in a $Ar-O_{2}$ mixture to form the oxide
barrier of the junctions. The chip is glued with wax on a printed
circuit board (PCB) and wire bonded to it. The PCB is then screwed
in a copper box anchored to the cold plate of a dilution refrigerator.\\

\paragraph{S2. Sample parameters\protect \\
}

The sample is first characterized by spectroscopy (see Fig.$\,$1.b
of main text). The incident power used is high enough to observe the
resonator frequency $\nu_{\mathrm{R}}$, the qubit line $\nu_{01}$,
and the two-photon transition at frequency $\nu_{02}/2$ between the
ground and second excited states of each transmon (data not shown).
A fit of the transmon model to the data yields the sample parameters
$E_{\mathrm{J}}^{\mathrm{I}}/h=36.2\,\mathrm{GHz}$, $E_{\mathrm{C}}^{\mathrm{I}}/h=0.98\,\mathrm{GHz}$,
$d_{I}=0.2$, $E_{\mathrm{J}}^{\mathrm{II}}/h=43.1\,\mathrm{GHz}$,
$E_{\mathrm{C}}^{\mathrm{II}}/h=0.87\,\mathrm{GHz}$, $d_{\mathrm{II}}=0.35$,
$\nu_{\mathrm{R}}^{\mathrm{I}}=6.84\,\mathrm{GHz}$, and $\nu_{\mathrm{R}}^{\mathrm{II}}=6.70\,\mathrm{GHz}$.
The qubit-readout anticrossing at $\nu=\nu_{\mathrm{R}}$ yields the
qubit-readout couplings $g_{0}^{\mathrm{I}}\simeq g_{0}^{\mathrm{II}}\simeq(2\pi)\,50\,\mathrm{MHz}$.
Independent measurements of the resonator dynamics (data not shown)
yield quality factors $Q_{\mathrm{I}}=Q_{\mathrm{II}}=730$ and Kerr
non linearities {[}13,\cite{FlorianKerr}{]} $K_{\mathrm{I}}/\nu_{\mathrm{R}}^{\mathrm{I}}\simeq K_{\mathrm{II}}/\mathrm{\nu}_{\mathrm{R}}^{II}\simeq-2.3\pm0.5\times10^{-5}$.\\

\paragraph{S3. Experimental setup}
\begin{itemize}
\item Qubit resonant microwave pulses: The qubit drive pulses are generated
by two phase-locked microwave generators feeding a pair of I/Q-mixers.
The IF inputs are provided by a 4-Channel$1\,\mathrm{GS/s}$ arbitrary
waveform generator (AWG Tektronix AWG5014). Single-sideband mixing
in the frequency range of 50-300 MHz is used to generate multi-tone
drive pulses and to obtain a high ON/OFF ratio ($>\,50\,\mathrm{dB}$).
Phase and amplitude errors are corrected by applying suitable sideband
and carrier frequency dependent corrections to the amplitude and offset
of the IF signals. 
\item Qubit frequency control: Flux control pulses are generated by a second
AWG and sent to the chip through a transmission line equipped with
40$\,$dB total attenuation and a pair of 1 GHz dissipative low-pass
filters at $4\,\mathrm{K}$. The input signal of each flux line is
returned to room temperature through an identical transmission line
and measured, which allows to compensate the non-ideal frequency response
of the line.
\item Readout pulses: The driving pulses for the Josephson bifurcation amplifier
(JBA) readouts are generated by mixing the continuous signals of a
pair of microwave generators with IF pulses provided by a $1\,\mathrm{GS/s}$
arbitrary waveform generator (AWG Tektronix AWG5014). Each readout
pulse consists of a measurement part with a rise time of $30\,\mathrm{ns}$
and a hold time of 100 ns, followed by a $2\,\mu s$ long latching
part at 90 \% of the pulse height. 
\item Drive and measurement lines: The drive and readout microwave signals
of each qubit are combined and sent to the sample through a pair of
transmission lines with total attenuation 70 dB and filtered at $4\mathrm{\, K}$
and $300\,\mathrm{mK}$. A microwave circulator at $20\,\mathrm{mK}$
protects the chip from the amplifier noise. The signals are amplified
by $36\,\mathrm{dB}$ at $4\,\mathrm{K}$ by two cryogenic HEMT amplifiers
(CIT Cryo 1) with noise temperature $5\,\mathrm{K}$. The reflected
readout pulses are amplified and demodulated at room temperature.
The IQ quadratures of the demodulated signals are sampled at $1\,\mathrm{GS/s}$
by a 4-channel data acquisition system (Acqiris DC282). 
\end{itemize}

\paragraph{S4. Readout Errors\protect \\
}

\begin{figure}
\includegraphics[width=8cm]{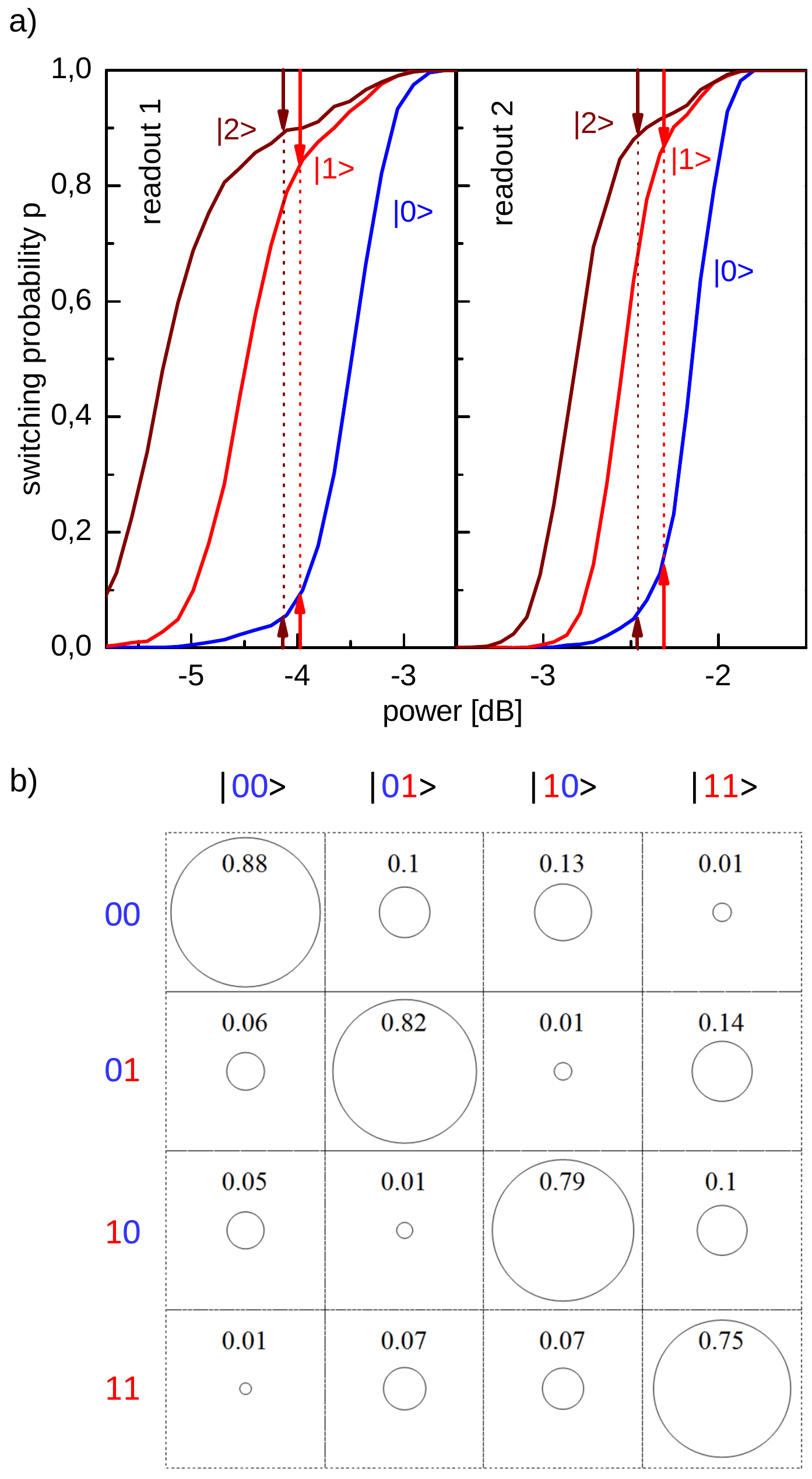} \caption{\label{fig:readouterrors} }
(a) Switching probability $p$ of each readout as a function of its
peak driving power, when its qubit is prepared in state $\left|0\right\rangle $
(blue), $\left|1\right\rangle $ (red), or $\left|2\right\rangle $
(brown), with the other qubit being far detuned. The arrows indicate
the readout errors where the contrast is optimal with (brown) and
without (red) $\left|1\right\rangle \rightarrow\left|2\right\rangle $
shelving. (b) Readout matrix giving the probabilities of the four
$ab$ outcomes, for the four computational input states $|uv\rangle$,
when using $\left|1\right\rangle \rightarrow\left|2\right\rangle $
shelving. %
\end{figure}

Errors in our readout scheme are discussed in detail in Ref. {[}13{]}
for a single qubit. First, incorrect mapping $\left|0\right\rangle \rightarrow1$
or$\left|1\right\rangle \rightarrow0$ of the projected state of the
qubit to the dynamical state of the resonator can occur, due to the
stochastic nature of the switching between the two dynamical states.
As shown in Fig. S4.1, the probability $p$ to obtain the outcome
1 varies continuously from 0 to 1 over a certain range of drive power
$P_{\mathrm{d}}$ applied to the readout. When the shift in power
between the two $p_{\left|0\right\rangle ,\left|1\right\rangle }(P_{\mathrm{d}})$
curves is not much larger than this range, the two curves overlap
and errors are significant even at the optimal drive power where the
difference in $p$ is maximum. Second, even in the case of non overlapping
$p_{\left|0\right\rangle ,\left|1\right\rangle }(P_{\mathrm{d}})$
curves, the qubit initially projected in state$\left|1\right\rangle $
can relax down to $\left|0\right\rangle $ before the end of the measurement,
yielding an outcome 0 instead of 1. The probability of these two types
of errors vary in opposite directions as a function of the frequency
detuning $\Delta=\nu_{\mathrm{R}}-\nu>0$ between the resonator and
the qubit, so that a compromise has to be found for $\Delta$. As
explained in the main text, we use a shelving method to the second
excited state in order to improve the readout contrast $c=Max\left(p_{\left|1\right\rangle }-p_{\left|0\right\rangle }\right)$,
with a microwave $\pi$ pulse at frequency $\nu_{12}$ bringing state
$\left|1\right\rangle $ into state $\left|2\right\rangle $ just
before the readout pulse. The smallest errors $e_{0}^{\mathrm{I,II}}$
and $e_{1}^{\mathrm{I,II}}$ when reading $\left|0\right\rangle $
and $\left|1\right\rangle $ are found for $\Delta_{\mathrm{I}}=440\,\mathrm{MHz}$
and $\Delta_{\mathrm{II}}=575\,\mathrm{MHz}$: $e_{0}^{\mathrm{I}}=5\%$
and $e_{1}^{\mathrm{I}}=13\%$ (contrast $c_{\mathrm{I}}=1-e_{0}^{\mathrm{I}}-e_{1}^{\mathrm{I}}=82\%$),
and $e_{0}^{\mathrm{II}}=5.5\%$ and $e_{1}^{\mathrm{II}}=12\%$ ($c_{\mathrm{II}}=82\%$).
When using the $\left|1\right\rangle \rightarrow\left|2\right\rangle $
shelving before readout, $e_{0}^{\mathrm{I}}=2.5\%$ and $e_{2}^{\mathrm{I}}=9.5\%$
(contrast $c_{\mathrm{I}}==1-e_{0}^{\mathrm{I}}-e_{2}^{\mathrm{I}}=88\%$),
and $e_{0}^{\mathrm{II}}=3\%$ and $e_{2}^{\mathrm{II}}=8\%$ ($c_{\mathrm{II}}=89\%$).
These best results are very close to those obtained in {[}12{]}, but
cannot however be exploited for simultaneous readout of the two qubits.

Indeed, when the two qubits are measured simultaneously, we find an
influence of the projected state of each qubit on the outcome of the
readout of the other one. In order to to minimize this spurious effect,
we increase the detuning $\Delta_{\mathrm{I,II}}$ up to $\sim1\,\mathrm{GHz}$
with respect to previous optimal values. An immediate consequence
shown in Fig. S4.1(a) is a reduction of the $c_{\mathrm{I,II}}$ contrasts.
The errors when reading $\left|0\right\rangle $ and $\left|1\right\rangle $
are then $e_{0}^{\mathrm{I}}=10\,\%$ and $e_{1}^{\mathrm{I}}=16\,\%$
(contrast $c_{\mathrm{I}}=74\%$) and $e_{0}^{\mathrm{II}}=12\,\%$
and $e_{1}^{\mathrm{II}}=15\,\%$ (contrast $c_{\mathrm{II}}=73\%$).
When shelving the qubit in state $\left|2\right\rangle $ , the errors
are $e_{0}^{\mathrm{I}}=5\,\%$, $e_{2}^{\mathrm{I}}=11\,\%$ (contrast
$c_{\mathrm{I}}=84\%$), $e_{0}^{\mathrm{II}}=5\,\%$, $e_{2}^{\mathrm{II}}=12\,\%$
(contrast $c_{\mathrm{I}}=83\%$). The readout errors are captured
in the $4\times4$ readout matrix $\mathcal{R}$ shown in Fig. S4.1(c),
that gives the probabilities $p_{\mathrm{uv}}$ of the four possible
outcomes for the different input states using the $\left|1\right\rangle \rightarrow\left|2\right\rangle $
shelving technique. This matrix $\mathcal{R}$ is used to correct
the readout errors only when doing state tomography, and not when
the running the algorithm once. The cause of the readout crosstalk
in our processor is discussed in Ref. {[}11{]}.

\paragraph{S5. Algorithm Fidelity\protect \\
}

The fidelity of each possible outcome $ab$ $\in\{00,01,10,11\}$
of our algorithm is given as \[
f_{\mathrm{ab}}=p_{\mathrm{ab/|ab\rangle}}/\left(p_{\mathrm{ab/|00\rangle}}+p_{\mathrm{ab/|01\rangle}}+p_{\mathrm{ab/|10\rangle}}+p_{\mathrm{ab/|11\rangle}}\right),\]
where $p_{\mathrm{ab/|uv\rangle}}$ is the conditional probability
for obtaining $ab$ when the state $|uv\rangle$ has been marked by
the oracle $O_{\mathrm{uv}}$. Table \ref{tab:Probabilities-for-obtaining}
shows these probabilities $p_{\mathrm{ab/|uv\rangle}}$ for all possible
combinations of $ab$ and $|uv\rangle$ as well as the fidelities
$f_{\mathrm{ab}}$. The average fidelity of the algorithm is 59.1
\%.

\begin{table}[H]
\begin{centering}
\begin{tabular}{|c|c|c|c|c|c|c|}
\hline 
$ab$/$|uv\rangle$ & $\left|00\right\rangle $ & $\left|01\right\rangle $ & $\left|10\right\rangle $ & $\left|11\right\rangle $ & $\sum$ & $f_{ab}$\tabularnewline
\hline
\hline 
00 & 0.666 & 0.192 & 0.188 & 0.122 & 1.168 & 57.0 \%\tabularnewline
\hline 
01 & 0.127 & 0.554 & 0.071 & 0.122 & 0.874 & 63.4 \%\tabularnewline
\hline 
10 & 0.128 & 0.106 & 0.615 & 0.239 & 1.088 & 56.5 \%\tabularnewline
\hline 
11 & 0.079 & 0.148 & 0.126 & 0.517 & 0.870 & 59.4 \%\tabularnewline
\hline
\end{tabular}
\par\end{centering}

\caption{\label{tab:Probabilities-for-obtaining}Conditional probabilities
$p_{ab/|uv\rangle}$ and statistical fidelities $f_{ab}$ for all
possible outcomes $ab$, measured for our version of Grover's algorithm.}
\end{table}

\end{document}